\documentstyle[epsfig,preprint,aps,tighten]{revtex}

\begin{document}
\title{Loop Quantum Mechanics and the Fractal Structure of
Quantum Spacetime}

\author{Stefano Ansoldi}
\address{Dipartimento di Fisica Teorica\break Universit\`a di Trieste,\\
INFN, Sezione di Trieste\\
e-mail: ansoldi@trieste.infn.it}

\author{Antonio Aurilia} \address{Department of Physics\\
California State Polytechnic University\break Pomona, CA
91768\\
e-mail:aaurilia@csupomona.edu}

\author{Euro Spallucci}
\address{Dipartimento di Fisica Teorica\break Universit\`a di
Trieste,\\
INFN, Sezione di Trieste\\
   e-mail: spallucci@trieste.infn.it }
\maketitle

\begin{abstract}
We discuss the relation between string quantization based on
the Schild path integral and the Nambu--Goto path integral.
The equivalence between the two approaches at the classical
level is extended to the quantum level by
a saddle--point evaluation of the corresponding path
integrals. A possible relationship between $M$--Theory and
the quantum mechanics of string loops is pointed out. Then,
within the framework of ``loop quantum mechanics'', we
confront the difficult question as to what exactly gives rise
to the structure of spacetime. We argue that the large scale
properties of the string condensate are responsible for the
effective Riemannian geometry of classical spacetime. On the other hand, 
near the Planck scale the condensate ``evaporates'', and what
is left behind is a  ``vacuum'' characterized by an effective {\it fractal}
geometry.
\end{abstract}
\vfill
\noindent
{\small Invited paper to appear in the special issue
of {\it Chaos, Solitons and Fractals} on ``Super strings,
M,F,S,...Theory'' (M.S. El Naschie and C.Castro, eds.)}
\newpage
\newcommand\beq{\begin{equation}}
\newcommand\eeq{\end{equation}}

\newcommand\beqa{\begin{eqnarray}}
\newcommand\eeqa{\end{eqnarray}}

\bigskip

\section{Introduction}
\begin{figure}
\begin{center}
\hskip -3cm \psfig{figure=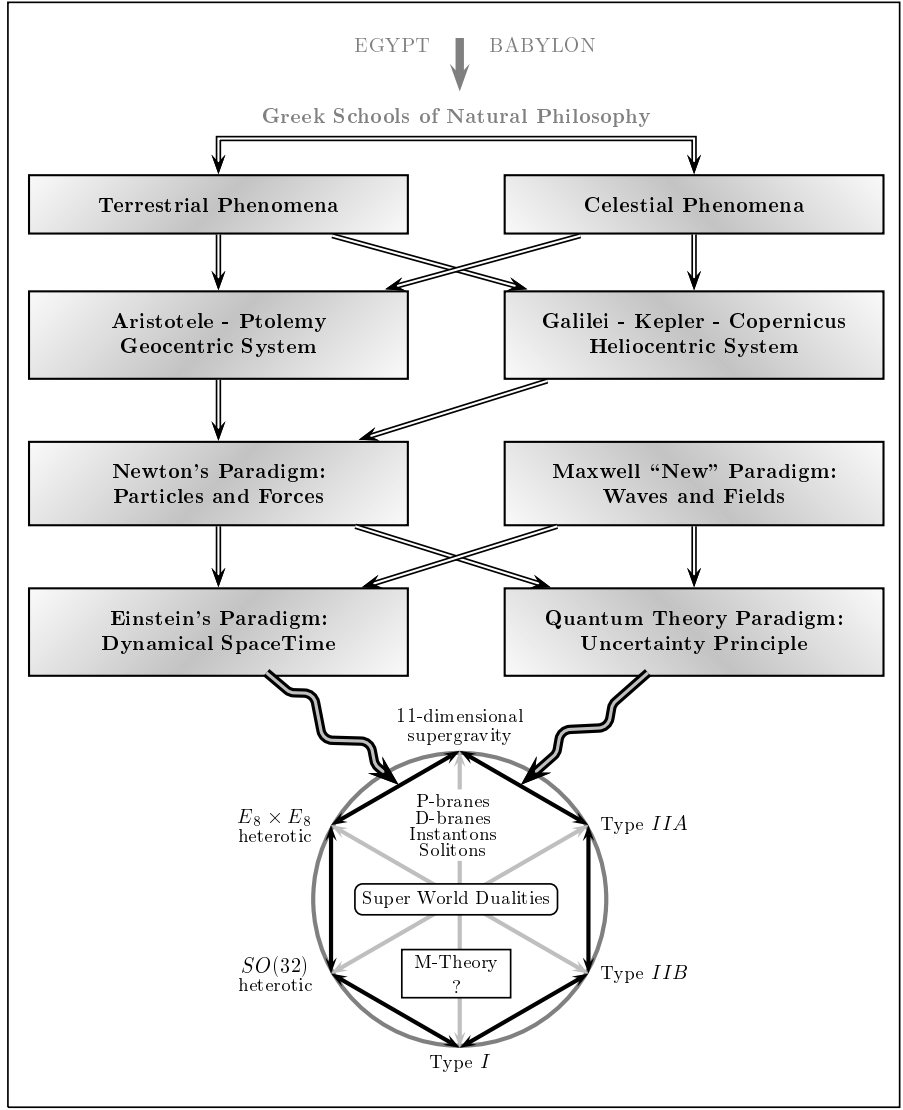,width=15cm}
\end{center}
\caption{History of physics shows that conflicting theories eventually
merge into a broader and deeper synthesis. Will M-Theory lead to a UNIQUE
supersynthesis of quantum theory, gravity theory and supersymmetry?}
\label{fig}
\end{figure}

At its most fundamental level, research in theoretical high
energy physics means research about the nature of mass and
energy, and ultimately about the structure of space and time.
It may even be argued that the whole history of physics, to a
large extent, represents the history of the ever changing
notion of space and time in response to our ability to probe
infinitesimally small distance scales as well as larger and
larger cosmological distances.\\
The ``flow chart'' in Figure \ref{fig} summarizes the dialectic
process which has led, through nearly twenty five hundred
years of philosophical speculation and scientific inquiry,
to the current theoretical efforts in search of a
supersynthesis of the two  conflicting paradigms of 20--th
century physics, namely, the Theory of General Relativity and
Quantum Theory. In that Hegelian perspective of the history
of physics, such a supersynthesis is regarded by many as the
``holy grail'' of contemporary high energy physics. However,
the story of the many efforts towards the formulation of that
synthesis, from supergravity to superbranes, constitutes, in
itself, a fascinating page in the history of theoretical
physics at the threshold of the new millennium. The early
`80s excitement about string theory (``{\it{}The
First String Revolution}'') followed from the prediction that
only the gauge groups $SO(32)$ and $E(8)\otimes E(8)$
provide a quantum mechanically consistent, i.e.,
anomaly free, unified theory which includes gravity\cite{gs},
and yet is capable, at least in principle, of
reproducing the standard electro--weak theory below the GUT
scale. However, several fundamental questions were left
unanswered. Perhaps, the most prominent one regards the
choice of the compactification scheme required to bridge the gap between
the multi--dimensional, near--Planckian string--world, and
the low energy, four dimensional universe we live in \cite{nara}. Some related
problems, such as the vanishing of the cosmological constant (is it
really vanishing, after all?) and the breaking of supersymmetry
were also left without a satisfactory answer. The common feature of all
these unsolved problems is their intrinsically {\it non--perturbative}
character. More or less ten years after the
First String Revolution, the second one, which is still in
progress, has offered a second important clue into the nature
of the superworld. The diagram in Figure \ref{fig} encapsulates the
essential pieces of a vast mosaic out of which the final
theory of the superworld will eventually emerge. Among those
pieces,  the six surviving viable supermodels known at
present, initially thought to be candidates for the role of a
fundamental Theory of Everything, are now regarded as
different asymptotic realizations, linked by a web of
dualities, of a unique and fundamentally new paradigm of
physics which goes under the name of {\it M--Theory}\cite{schwarz}.
The essential components of this
underlying matrix theory appear to be string--like objects as
well as other types of extendons, e.g.,
P--branes, D--branes, ..., (any letter)--branes. Moreover, a
new computational approach is taking shape which is based on
the idea of trading off the strongly coupled regime of a
supermodel with the weakly coupled regime of a different
model through a systematic use of dualities. \\
 Having said that, the fact remains that M--theory, at
present, is little more than a name for a mysterious
supertheory yet to be fully formulated. In particular, we
have no clue as to what radical modification it will bring to
the notion of spacetime in the short distance regime. In the
meantime, it seems reasonable to attempt to isolate the
essential elements of such non--perturbative approach to the
dynamics of extended objects. One such approach that we have
developed over the last few years \cite{vanzetta},\cite{noi}, \cite{noi3},
is a refinement of an early formulation of {\it quantum string theory} by
T.Eguchi \cite{eguchi}, elaborated by following a formal analogy with a
Jacobi--type formulation of the {\it canonical quantization of gravity}. \\
Thus, our immediate objective, in the following Section, is
to illustrate the precise meaning of that analogy. In
Sections III and IV we discuss our quantum mechanical
elaboration of Eguchi's approach in terms of ``areal'' string
variables, string propagators and string wave functionals.
This discussion, which can be easily extended to p--branes of
higher dimensionality, enables us to exemplify a possible
relationship between M--theory and the quantum mechanics of
string loops. Section V is divided into two subsections where
we discuss the functional Schr\"odinger equation of ``loop
quantum mechanics'' and its solutions in order to derive the
Uncertainty Principle for strings as well as its principal
consequence, namely, the fractalization of quantum spacetime.
We then conclude our discussion of the structure of spacetime
in terms of an effective lagrangian based on a covariant,
functional extension of the Ginzburg--Landau model of
superconductivity.

\section{Gravity/String Quantization Schemes}

There is an intriguing similarity between the problem of
quantizing gravity, as described by General Relativity, and
that of
quantizing a relativistic string, or any higher dimensional
relativistic extended object. In either case, one can follow
one of two main routes: i) a quantum field theory
inspired--{\it covariant quantization},
or ii) a {\it canonical quantization} of the Schr\"odingertype.
The basic idea underlying the covariant approach is to
consider the metric tensor $g_{\mu\nu}(x)$ as an ordinary
matter field and follow the standard quantization procedure,
namely, Fourier analyze small fluctuation around a classical
background configuration and give the Fourier coefficients
the
meaning of creation/annihilation operators of the
gravitational field quanta, the {\it gravitons}. In the
same fashion, quantization of
the string world--sheet fluctuations leads to a whole
spectrum of particles with different values of mass and spin. Put briefly,

\begin{eqnarray}
&& g_{\mu\nu}(x)=\hbox{background}\,+\,\hbox{``graviton''}\nonumber\\
&& X^{\mu}(\tau,s)=\hbox{zero--mode}\,+\,\hbox{particle spectrum}\nonumber
\quad .
\end{eqnarray}

Against this background, one may elect to forgo the full
covariance
of the quantum theory of gravity in favor of the more
restricted
symmetry under transformations preserving the ``{\it
canonical spacetime slicing}'' into a one--parameter family
of space-like three--surfaces. This splitting of space and
time amounts to selecting the spatial components of the
metric, modulo three--space reparametrizations,
as the gravitational degrees of freedom to be quantized. This
approach focuses on the quantum mechanical description of the
space itself, rather than the corpuscular content of the
gravitational field. Then, the quantum state of the spatial
3--geometry is controlled by
the Wheeler--DeWitt equation \begin{equation}
\left[
\hbox{Wheeler--DeWitt operator}\right]\Psi[G_3]=0\label{wd}
\end{equation}
and the wave functional $\Psi[G_3]$, {\it the wave function
of the universe}, assigns a probability amplitude to each
allowed three
geometry. The relation between the two quantization
schemes is akin to the relationship in particle dynamics
between
{\it first quantization}, formulated in terms of single
particle
wave functions along with the corresponding Schr\"odinger
equation,  and the {\it second quantization} expressed in 
terms of
creation/annihilation operators along with the corresponding
field equations. 
Thus, covariant quantum gravity is, {\it conceptually}, a 
second 
quantization framework for calculating amplitudes, cross 
sections, 
mean life,etc., for any physical process involving graviton 
exchange.
Canonical quantum gravity, on the other hand, is a 
Schr\"odinger--type
first quantization framework, which assigns a probability
amplitude
for any allowed geometric configuration of three dimensional
physical space. It must be emphasized that there is no
immediate
relationship between the graviton field and the wave function
of the universe. Indeed, even if one elevates
$\Psi[G_3]$ to the role of {\it field operator,} it would
create or destroy  {\it
entire three surfaces} instead of single gravitons. In a more
pictorial  language, the
wave functional $\Psi[G_3]$ becomes a quantum operator
creating/destroying full universes! Thus, in order to avoid
confusion with quantum gravity as the ``theory of gravitons''
, the quantum field theory of universes is referred to as the
{\it third quantization scheme}, and has been investigated
some years ago mostly in connection with the cosmological
constant problem. \\
Of course, as far as gravity is concerned, any quantization
scheme is affected
by severe problems: perturbative covariant quantization of
General
Relativity is not renormalizable, while the
intrinsically non--perturbative Wheeler--DeWitt equations can
be solved only under extreme simplification such as the mini--superspace
approximation. These shortcomings provided the
impetus toward the formulation of superstring theory as the
only consistent  quantization scheme which accommodates the
graviton in its (second quantized) particle spectrum. Thus,
according to the prevalent way of thinking, there is no
compelling reason, nor clear cut procedure to formulate a
first quantized quantum
mechanics of relativistic extended objects. In the case of
strings, this attitude is deeply rooted in the conventional
interpretation of the world--sheet coordinates
$X^{\mu}(\tau,s)$ as a ``multiplet of
scalar fields'' defined over a two--dimensional manifold
covered by the $\{\tau,s \}$ coordinate mesh. According to
this point of view,
quantizing a relativistic string is formally equivalent to
quantizing
a two--dimensional field theory, bypassing a preliminary
quantum mechanical formulation. However, there are at least 
two 
objections against this kind of reasoning. The first follows 
from 
the analogy between the canonical formulation of General 
Relativity and 3--brane dynamics, and the second objection 
follows from the
``Schr\"odinger representation'' of quantum field theory. More
specifically:
i) the Wheeler-- DeWitt equation can be interpreted, in a
modern
perspective, as the wave equation for the orbit of a
relativistic 3--brane. In this perspective, then, why not conceive of a
similar equation for a 1--brane? ii) the functional
Schr\"odinger representation of quantum field theory assigns a
probability amplitude to each field configuration over a
space--like slice $t=const.$, and the corresponding wave
function obeys a functional Schr\"odinger--type equation.\\
Pushing the above arguments to their natural conclusion, we
are led to entertaining the interesting possibility of
formulating a {\it functional}
quantum mechanics for strings and other p--branes. This
approach has received scant attention in the mainstream work
about quantum string
theory, presumably because it requires an explicit breaking
of the celebrated reparametrization invariance, which is
the distinctive symmetry of relativistic extended objects.\\
All of the above reasoning leads us to the central question
that we wish to analyze, namely: is there any way to
formulate a reparametrization
invariant string quantum mechanics?\\
As a matter of fact, a possible answer was suggested by T.
Eguchi as early as 1980
\cite{eguchi}, and our own elaboration of that quantization
scheme \cite{noi} is the topic of Section III.

\section{Eguchi's Areal Quantization Scheme }

\subsection{The original formulation}
Eguchi's approach to string quantization follows closely  the
point--particle quantization along the guidelines of the
Feynman--Schwinger method. The essential point is that
reparametrization invariance is not assumed as an original
symmetry of the classical action;  rather, it is a symmetry
of the
physical Green functions to be obtained at the very end of 
the 
calculations by means of an averaging procedure over the 
string manifold 
parameters. More explicitly, the basic action is not the 
Nambu--Goto 
proper area of the string world--sheet, but the ``square'' of
it, i.e. the Schild Lagrangian \cite{schild}
\begin{equation}
L_{Schild}={1\over 4}\left[{\partial( x^\mu, x^\nu)\over
\partial(\tau,\sigma)}\right]^2\ ,\qquad {\partial( x^\mu,
x^\nu)\over \partial(\tau,\sigma)}\equiv \partial_\tau
x^\mu\wedge \partial_\sigma x^\nu  .
\end{equation}
The corresponding (Schild) action is invariant under area
preserving
transformations only, i.e.
\begin{equation}
(\tau,\sigma)\longrightarrow (\tau',\sigma'):
{\partial(\tau',\sigma') \over \partial(\tau,\sigma)}=1 .
\end{equation}
Such a restricted symmetry requirement leads to a
new, Jacobi--type, canonical formalism in which {\it the
world--sheet
proper area of the string manifold plays the role of
evolution parameter}. In other words, the ``proper time'' is
neither
$\tau$ or $x^0$, but the {\it invariant} combination of
target and internal space coordinates
$x^\mu$ and $\sigma^a=(\tau,\sigma)$ provided by
\begin{equation}
A=\int d^2\sigma\sqrt{-\gamma}\ ,\quad
\gamma\equiv\det{\partial_a x^\mu \partial_b x_\mu} .
\label{Adef}
\end{equation}
Once committed to this unconventional definition of time, the
quantum amplitude for the transition from an initial
vanishing configuration to a final non--vanishing string
configuration
after a lapse of areal time $A$, is provided by the kernel
$G(x(s);A)$ which satisfies the following  diffusion--like
equation, or imaginary
area Schr\"odinger equation
\begin{equation}
{1\over 2}{\delta^2\over \delta x^\mu(s) \delta
x_\mu(s)}G(x(s);A)= {\partial\over\partial A}G(x(s);A)  .
\end{equation}
Here, $x^\mu(s)=x^\mu(\tau(s), \sigma(s))$ represents
the physical string coordinate, i.e. the only space-like boundary
of the world--sheet of area $A$. It may be worth emphasizing
at this point that in quantum
mechanics of point particles the ``time'' $t$
is {\it not a measurable quantity
but an arbitrary parameter}, since there does not exist a
self--adjoint quantum operator with eigenvalues $t$.
Similarly, since there is no self--adjoint operator
corresponding to
the world--sheet area, $G(x(s);A)$ turns out to be explicitly
dependent on the arbitrary parameter $A$, and cannot have an
immediate
physical meaning. However, the Laplace transformed Green
function
is $A$--independent and corresponds to the Feynman
propagator
\begin{eqnarray}
G(x(s);M^2)&\equiv &\int_0^\infty dA G(x(s);A) \exp(-M^2
A/2)\nonumber\\
&=&-{1\over 2(2\pi)^{3/2}}\int {dA\over A^{3/2}} \exp\left(-
{F\over 2A}-{1\over 2}M^2A\right)\\ F&=&{1\over
4}\left(F^{\mu\nu}\pm {}^*F^{\mu\nu}\right)^2\ ,\quad
F^{\mu\nu}[C]=\int_C x^\mu dx^\nu
\end{eqnarray}
where $F$ stands for the self--dual (anti self--dual) area
element.\\
Evidently, this approach is quite different  from the
``normal
mode quantization'' based on the Nambu--Goto action or the
path integral formulation a la' Polyakov, and our immediate
purpose, in the next subsection, is to establish a connection
with the conventional
path integral quantization of a relativistic string. Later,
in Section IV, we speculate about a possible connection
between our functional approach and a recently formulated
matrix model of Type IIB superstring.

\subsection{Quantum Mechanics in Loop Space}

The Eguchi quantization program is essentially a sort of
quantum mechanics formulated in a space of string loops,
i.e., a space in which each point  represents a possible
geometrical configuration of
a closed string. To establish a connection with the
Nambu--Goto, or Polyakov path integral, it is advantageous to
start
from the quantum kernel
\begin{eqnarray}
&&K[C,C_0;A]=\int_{C_0}^C\int_{\gamma_0}^\gamma
D[\mu(\sigma)] \exp \left(iS[x,\xi,p,\pi,N]\right)\\
&&D[\mu(\sigma)]\equiv D[x(\sigma)] D[\xi(\sigma)]
D[P(\sigma)] D[\pi(\sigma)] D[N(\sigma)]
\end{eqnarray}
where the histories connecting the initial string
$C_0$ to the final one $C$ are weighted by the exponential of
the {\it reparametrized Schild action}
\begin{eqnarray}
S[X,P,\xi,\pi, N]=&&{1\over 2}\int_X\, P_{\mu\nu}\,
dx^\mu\wedge dx^\nu
+{1\over 2}\int_\Xi\, \pi_{ab}\, d\xi^a\wedge
d\xi^b\nonumber\\ -&&{1\over 2}\int_\Sigma
d^2\sigma\,N^{ab}(\sigma)\left[\pi_{ab} -\epsilon_{ab}
H(P)\right]  .
\end{eqnarray}
The ``dictionary'' used in the above equation is as follows:
i) $P_{\mu\nu}$ is the momentum canonically conjugated
to the world--sheet area element; ii) $ H(P)=P_{\mu\nu}
P_{\mu\nu}/4m^4 $ is the Schild Hamiltonian density, and iii)
$m^2=1/4\pi\alpha'$ is the string tension. Furthermore,
the original world--sheet coordinates $\xi^a$ have been
promoted to the role of {\it dynamical variables,} i.e., they
now represent
fields $\xi^a(\sigma)$ defined over the string manifold, $$
\xi^a\longrightarrow \xi^a(\sigma^m)\ , \quad m=0,1 $$
and $\pi _{ab}$, the momentum conjugate to $\xi^a $, has been introduced
into the Hamiltonian form of the action. The relevant
dynamical quantities in loop space are listed in
Table~[\ref{uno}].\\ The enlargement of the canonical
phase space endows the Schild action with the full
reparametrization invariance under the transformation
$\sigma^m\longrightarrow\zeta^m(\sigma)$, while {\it
preserving the polynomial structure} in the dynamical
variables, which is a necessary condition to solve the path
integral. The regained reparametrization invariance forces
the new (extended) hamiltonian to be weakly vanishing, i.e.,
$H(P)-
\epsilon^{ab}\pi_{ab}/2\approx 0$. The quantum implementation
of this condition is carried out by means of the Lagrange
multiplier
$N^{ab}(\sigma)$.\\
By integrating out $\pi^{ab}$ and $\xi^a$ one obtains
\begin{eqnarray}
K[C,C_0;A]&=&\int_0^\infty d{\cal E} e^{i{\cal E}A}
\int_{C_0}^C D[x(\sigma)] D[P(\sigma)]
D[N(\sigma)]\times\nonumber\\ &&
\qquad \times \exp\left\{{i\over 2}\int_X\,
P_{\mu\nu}\,
dx^\mu\wedge dx^\nu-{i\over 2}\epsilon_{ab}\int_\Sigma
d^2\sigma\,N^{ab}(\sigma)\left[{\cal E}
- H(P)\right]\right\}\\
&\equiv& 2i m^2 \int_0^\infty d{\cal E} e^{i{\cal E}A}
G[C,C_0;{\cal E}]\label{fourier}
\end{eqnarray}
\begin{eqnarray}
G[C,C_0;{\cal E}]&=&
\int_{C_0}^C D[x]D[N]\exp\left\{
-i\int_\Sigma d^2\sigma \left[
-{m^2\over 4N(\sigma)} \dot{x}^{\mu\nu} \dot{x}_{\mu\nu} + N(\sigma) {\cal
E} \right]\right\}\nonumber\\
&=&\int_{C_0}^C D[x(\sigma)]D[N(\sigma)]\exp\left\{-i
S_{Schild}[x,N]\right\}\label{prop}  .
\end{eqnarray}
The above expressions show the explicit relation between the
{\it fixed area string 
propagator } $K[C,C_0;A]$, and the {\it fixed ``energy''
string propagator} $G[C,C_0;{\cal E}]$ without recourse to
any {\it ad hoc}
averaging prescription in order to eliminate the $A$
parameter
dependence. Moreover, the {\it saddle point} value of the
string propagator
(\ref{prop}) is evaluated to be
\begin{equation}
G[C,C_0;{\cal E}]\simeq
\int_{C_0}^C D[x(\sigma)]\exp\left\{
-i\sqrt{m^2{\cal E}}\int_\Sigma d^2\sigma
\sqrt{-\dot{x}^{\mu\nu} \dot{x}_{\mu\nu}}\right\}\label{ng} .
\end{equation}
Since ${\cal E}$ has dimension of inverse length square, in
natural units, we can set the
string tension equal to $m^2$, and then (\ref{ng}) reproduces
exactly the Nambu--Goto path integral.
This result allows us to establish the
following facts:\\ i)
Eguchi's approach corresponds to  quantizing a string by
keeping fixed {\it the area} of the string histories
in the path integral, and then taking the average over the
string tension
values;\\
ii) the Nambu--Goto approach, on the other hand, corresponds
to  quantizing a string
by keeping  fixed {\it the string tension } and then taking
the average over the world--sheet areas;\\
iii) the two quantization schemes are equivalent in the
saddle point approximation.\\
Finally, since the Schild propagator $K[C,C_0;A]$ can be computed exactly
\begin{equation}
K[C,C_0;A]=\left({m^2\over 2i\pi A}\right)^{3/2} \exp\left[
{im^2\over 4 A}\left(\sigma^{\mu\nu}(C)-\sigma^{\mu\nu}(C_0)
\right)^2\right]\ ,\quad
\sigma^{\mu\nu}(C)\equiv \oint_C x^\mu dx^\nu\label{exact}
\end{equation}
we obtain through Eqs.(\ref{fourier}), (\ref{prop})
a {\it non--perturbative} definition of the Nambu--Goto
propagator (\ref{ng}):
\begin{equation}
G[C,C_0;m^2]={1\over 2im^2}\int_0^\infty dA\, e^{-im^2A}\,
K[C,C_0;A] \end{equation}
where $K[C,C_0;A]$ is given by Eq.(\ref{exact}).

\section{Pl\"ucker coordinates and M--theory}

One of the most enlightening features of the Eguchi approach is
the formal correspondence it establishes between the quantum
mechanics of point--particles and  string loops. Such a relationship
is summarized in the {\it translation code} displayed in 
Table [\ref{dictionary}]. Instrumental to this correspondence
is the replacement of the canonical
string coordinates $x^\mu(s)$ with the reparametrization
invariant area elements $\sigma^{\mu\nu}[C]$. We shall refer
to these areal variables as Pl\"ucker coordinates\cite{pluck}.
Surprising as it may appear at first sight, the new
coordinates $\sigma^{\mu\nu}[C]$ are just as ``natural'' as
the old $x^\mu(s)$ for the purpose of  defining the string
``position''. A naive argument to support this claim goes as
follows. In the  Jacobi-- type formulation of particle
dynamics, the position of the physical object is provided by the instantaneous
end--point of its own world--line
\begin{equation}
x^\mu(P)\equiv x^\mu(T)=\int_{-\infty}^T d\tau {dx^\mu\over
d\tau } =\hbox{world--line end point}.
\end{equation}
In other words, the instantaneous position of the particle is given by the
line integral of the {\it tangent vector} up to the chosen final value
$T$. Similarly, then, it seems natural to define the ``string
position''
as the surface integral of the {\it tangent bi--vector} up to
the final boundary of the world--sheet
\begin{equation}
\sigma^{\mu\nu}[C]=\int_0^{s_0}d\sigma
\int_{-\infty}^T d\tau\,
\partial_\tau x^\mu\wedge \partial_\sigma x^\nu=
\hbox{world--surface boundary}
\label{sigmamunudef}
\end{equation}
which is nothing but the loop area element appearing in
Eq.(\ref{exact}). A geometric interpretation of the new
``matrix''--coordinate assignement to the loop $C$ is that
they represent {\it the areas of the loop projected shadows
onto the coordinate planes}. In this connection, note that the
$\sigma$--tensor satisfies the constraint
\begin{equation}
\epsilon_{\lambda\mu\nu\rho}\,\sigma^{\lambda\mu}\sigma^{\nu\rho}=0
\end{equation}
which ensures that there is a one--to--one correspondence 
between a given set of areas $\sigma^{\mu\nu}$ and a loop 
$C$. Thus, to summarize, the Pl\"ucker coordinates refer to
the {\it boundary} of the world sheet, and enter the string wave functional as an
appropriate set of position coordinates. Then, it is not
surprising that in such a formulation {\it homogeneity} 
requires a timelike coordinate with area dimension as
well\footnote{Note the
``Mach--ian flavor'' of this new definition of position and
time:
spacetime coordinates are not defined by themselves but only
in terms of objects located at a given point. This operative
definition of coordinates as labels of some material stuff
seems even more appropriate in  M--Theory
where {\it spacetime itself} is in
some way built out of fundamental strings, or branes, or
matrices, or... something else.}.\\
Finally, we note for the record, that the Pl\"ucker
coordinates provide a formal correspondence between string theory \cite{ka},  and a
certain class of electromagnetic field configurations
\cite{ngu}. As a matter of fact, a classical gauge field theory 
of relativistic strings was proposed several 
years ago \cite{no}, but only recently it was extended to
generic $p$--branes including their coupling to $p+1$--forms
and gravity \cite{noi2}.\\
Now that we have established the connection between areal
quantization and the path integral formulation of quantum 
strings, it seems almost compelling to ponder about the
relationship, if any, between the $\sigma^{\mu\nu}[C]$ matrix 
coordinates and the matrix coordinates which, presumably, lie 
at the heart of the M--Theory formulation of superstrings. 
Since the general framework of M--Theory is yet to be 
discovered, it seems reasonable to focus on a specific matrix 
model recently proposed for  {\it Type IIB} 
super strings\footnote{A similar matrix action for the Type 
IIA model has been
conjectured in \cite{banks}.}\cite{ikkt}. The dynamics of
this model is encoded 
into a simple Yang--Mills type action
\begin{equation}
S_{IKKT}=-{\alpha\over 4}\, Tr[\, A^\mu, A^\nu\,]^2 +\beta\,
Tr I +\hbox{fermionic part}\label{jap}
\end{equation}
where the $A^\mu$  variables are represented by $N\times N$
hermitian matrices, and $I$ is the unit matrix. {\it The
novelty of
this formulation is that it identifies the ordinary spacetime
coordinates with the eigenvalues of the non--commuting Yang--Mills
matrices}. In such a framework, the emergence of  {\it
classical spacetime} occurs in the {\it large--$N$
limit}, i.e., when the commutator goes into a Poisson
bracket\footnote{This is not the canonical Poisson bracket
which is replaced by the quantum mechanical commutator.
Rather, it is the world sheet symplectic structure which is
replaced by the (classical) matrix commutator for finite
$N$.}
and the matrix trace operation turns into a double continuous
sum over the row and column indices, which amounts to a two
dimensional
invariant integration. Put briefly,
\begin{eqnarray}
&&\lim _{N\rightarrow \infty}``Tr'' \rightarrow -i\int d\tau
d\sigma \sqrt{\gamma}\\
&&-i\lim _{N\rightarrow \infty}[\, A^\mu , A^\nu\,]
\rightarrow\{x^\mu ,x^\nu\}\\
&&\sqrt\gamma\{x^\mu ,x^\nu\}\equiv \dot{x}^{\mu\nu}\equiv \partial_\tau
x^\mu\wedge \partial_\sigma x^\nu\ .
\end{eqnarray}
What interests us is that, {\it in such a  classical limit,
the IKKT action}
(\ref{jap}) {\it turns into the Schild action in Eq.}(\ref{prop})
{\it once we make
the identifications}
\begin{equation}
\alpha\longleftrightarrow -m^2\ ,\quad
\beta \longleftrightarrow {\cal E}\ ,\quad
N(\tau,\sigma) \longleftrightarrow \sqrt{\gamma}  ,
\end{equation}
while the trace of the Yang--Mills commutator turns into the
oriented surface element
\begin{equation}
lim _{N\rightarrow \infty}Tr\left[ \,A^\mu, A^\nu
\,\right]\rightarrow \int_\Sigma d\tau d\sigma
\,\partial_\tau x^\mu\wedge \partial_\sigma x^\nu
\equiv \sigma^{\mu\nu}(\partial\Sigma )  .
\end{equation}
This formal relationship can be clarified by considering a
definite case. As an example let us consider a static $D$--string
 configuration both in the classical Schild formulation
and in the corresponding matrix description. It is
straightforward to prove that a length $L$ static string
stretched along the $x^1$ direction, i.e.
\begin{eqnarray}
&&x^\mu=\tau\,T \,\delta^{\mu\, 0}+{L\sigma\over
2\pi}\,\delta^{\mu\, 1}\ ,
\quad 0\le \tau \le 1\ ,\quad 0\le\sigma\le 2\pi\\ &&x^\mu=0\
,\quad \mu\ne 0,1
\end{eqnarray}
solves the classical equations of motion
\begin{equation}
\{x_\mu , \{x^\mu,x^\nu\}\, \}=0  .
\end{equation}
During a time lapse $T$ the string sweeps a time--like world--sheet
in the $(0,1)$--plane characterized by an area tensor
\begin{equation}
\sigma^{\mu\nu}(L,T)=\int_0^1 d\tau\int_0^{2\pi}
d\sigma\,\sqrt\gamma \,
\{x^\mu,x^\nu\}=TL\,\delta^{0[\,\mu}\delta^{\nu\,] 1}\ .\label{area}
\end{equation}
Eq.(\ref{area}) gives both the area and the orientation of
the rectangular loop  which is the boundary of the string
world--sheet. The
corresponding matrix solution, on the other hand, must
satisfy the equation
\begin{equation}
[\, A_\mu,[\, A^\mu,A^\nu\,]\,]=0\label{ym} \ .
\end{equation}
Consider, then, two hermitian, $N\times N$ matrices $\hat q$,
$\hat p$ with an approximate $c$--number commutation relation
$\displaystyle{ [\,\hat q,\hat p\,]= i}$, when $N>>1 $.
Then, a solution of the classical equation of motion (\ref
{ym}), corresponding to a {\it solitonic} state in string
theory, can be written as
\begin{equation}
A^\mu=T\,\delta^{\mu\, 0}\hat q+{L\over 2\pi}\,\delta^{\mu 1}\hat p \ .
\end{equation}
In the large--$N$ limit we find
\begin{equation}
-i[A^\mu,A^\nu]=-i {LT\over 
2\pi}\delta^{0[\,\mu}\delta^{\nu\,] 1}\, [\,\hat q,\hat p\,]\approx
{LT\over 2\pi}\delta^{0[\,\mu} \delta^{\nu\,] 1}
\end{equation}
and
\begin{equation}
-iTr[\, A^\mu,A^\nu\,]\approx\int_0^1
d\tau\int_0^{2\pi}\,\sqrt\gamma\, 
\{x^\mu,x^\nu\}=\sigma^{\mu\nu}(L,T)  .
\end{equation}

These results, specific as they are, point to a deeper
connection between the loop
space description of string dynamics and matrix models of
superstrings which, in our opinion,  
deserves a more detailed investigation. Presently, we shall 
limit ourselves to take a closer look at the functional 
quantum mechanics of string loops with an eye on its 
implications about the structure of spacetime in the short 
distance regime.

\section{Loop Quantum Mechanics}

\subsection{Correspondence Principle, Uncertainty Principle
and the Fractalization of Quantum Spacetime}

If history of physics is any guide, conflicting scientific
paradigms, as outlined in Figure \ref{fig}, generally lead to broader
and more predictive theories. Thus, one would expect that a
synthesis of general relativity and quantum theory will
provide, among other things, a deeper insight into the nature
and structure of space and time. Thus, reflecting on those
two major revolutions in physics of this century, Edward
Witten writes \cite{ed}, ``{\it Contemporary developments in
theoretical physics suggest that another revolution may be in
progress, through which a new source of ``fuzziness'' may
enter physics, and spacetime itself may be reinterpreted as
an approximate, derived concept.}''. \\
If spacetime is a derived concept, then is seems natural to
ask, ``what is the main property of the {\it fuzzy stuff},
let us call it {\it quantum spacetime}, that replaces the
smoothness of the classical spacetime manifold, and what is
the scale of distance at which the transition takes place?''.
Remarkably, the celebrated Planck length represents a very
near miss as far as the scale of distance is concerned. The
new source of fuzziness comes from string theory,
specifically from the introduction of the new fundamental
constant which determines the tension of the string.Thus, at
scales comparable to $(\alpha')^{1/2}$, spacetime becomes
fuzzy, even
in the absence of conventional quantum effects ($h=0$). While
the 
exact nature of this  fuzziness is unclear, it manifests 
itself in a new 
form of Heisenberg's principle, which now depends on both 
$\alpha'$ 
and $h$. Thus, in Witten's words, while ``{\it a proper
theoretical
framework for the [new] uncertainty principle has not yet
emerged,........the natural framework of the [string] theory
may
eventually prove to be inherently quantum mechanical.}''.\\
That new quantum mechanical framework may well constitute the
core of the yet undiscovered M--Theory, and the non
perturbative functional quantum mechanics of string loops
that we have developed in recent years may well represent a
first step on the long road toward a matrix formulation of
it. If this is the case, a challenging testing ground is
provided by the central issue of the structure of {\it
quantum spacetime}. This question was analyzed in Ref.\cite{noi3}
and we limit ourselves, in the remainder of
this subsection, to a brief elaboration of the arguments
presented there.\\
The main point to keep in mind, is the already mentioned 
analogy between ``loop quantum mechanics'' and the ordinary
quantum mechanics of point particles. That analogy is
especially evident in terms of the new {\it areal} variables,
namely, the spacelike area enclosed by the string loop, given
by Eq.(\ref{sigmamunudef}), and the timelike, proper area of the string
manifold, given by Eq.(\ref{Adef}). With that choice of dynamical
variables, the reparametrized formulation of the Schild
action principle
leads to the classical {\it energy per unit length
conservation}
$\displaystyle{H={\cal E}}$. Then, the {\it loop wave
equation} can
be immediately written down by translating this conservation
law in the quantum language through the {\it Correspondence
Principle}

\begin{eqnarray}
&&P_{\mu\nu}(s)\longrightarrow {i\over
\sqrt{x^{\prime\,2}(s)}} {\delta\over
\delta\sigma^{\mu\nu}(s)}\\ &&H\longrightarrow
-i{\partial\over\partial A}  .\end{eqnarray}
Thus, we obtain the Schr\"odinger equation anticipated in the
Introduction,
\begin{equation}
{1\over 4m^2l_C }\oint_C d\mu(s)
{\delta^2\Psi[\sigma\ ;A]\over \delta\sigma^{\mu\nu}(s)
\delta\sigma_{\mu\nu}(s)}=
i{\partial\over\partial A}\Psi[\sigma\ ;A]\label{ondauno}
\end{equation}
where we have introduced the {\it string wave functional}
$\Psi[\sigma\ ;A]$ as the amplitude to find the loop $C$
with area elements $\sigma^{\mu\nu}[C]$ as the only boundary
of a two--surface of internal area $A$. When no $A$--dependent
potential is present in loop space, the wave functional factors out as
\begin{equation}
\Psi[\sigma\ ;A]=\psi[\sigma]e^{-iA{\cal E}}
\end{equation}
and Eq.(\ref{ondauno}) takes the Wheeler--DeWitt form
\begin{equation}
{1\over 4}\oint_C d\mu(s)
{\delta^2\psi[\sigma]\over \delta\sigma^{\mu\nu}(s)
\delta\sigma_{\mu\nu}(s)}- m^2l_C{\cal E}\psi[\sigma]=0  .
\label{ondadue}
\end{equation}
Alternatively, one can exchange the area derivatives with the
more familiar functional variations of the string embedding
through the tangential projection
\begin{equation}
x^{\prime\,\mu}(s){\delta\over \delta\sigma^{\mu\nu}(s)}=
{\delta\over \delta x^\nu(s)}\ ,\quad
\end{equation}
where $x^{\prime\,\mu}(s)$ is the tangent vector to the
string loop. As a consistency check on our functional
equation, note that, if one further identifies the energy per
unit
length ${\cal E}$ with the string tension by setting
$\displaystyle{{\cal E}=m^2}$, then, Eq.(\ref{ondadue})
reads
\begin{equation}
{1\over l_C }\int_0^1 {ds\over\sqrt{x^{\prime\,2}}}
{\delta^2\psi[\sigma]\over \delta x^\mu(s) \delta
x_\mu(s)}=m^4\psi[\sigma]\label{ondatre} \end{equation}
which is the string field equation proposed several years
ago by Marshall and Ramond \cite{mr}. Note also that the
Schr\"odinger equation is a ``free'' wave
equation describing the random drifting of the string 
representative ``point'' in loop space. Perhaps, it is worth emphasizing
that this ``free motion'' in loop space is quite different from the free
motion of the string in physical space, not only because the physical string 
is  subject to its own tension, i.e. 
elastic forces are acting on it, but also because drifting  
from point to point in loop space corresponds physically to 
quantum mechanically 
jumping from string shape to string shape. Any  such ``shape
shifting'' process, random as it is, is subject to an extended
form of
the Uncertainty Principle which forbids the exact,
simultaneous
knowledge of the string shape and its area conjugate
momentum. The
main consequence of the Shape Uncertainty Principle is the
``fractalization'' of the string orbit in spacetime. The
degree of
fuzziness of the string world--sheet is measured by its
Hausdorff
dimension, whose {\it limiting value} we found to be
$\displaystyle{D_H=3}$. In order to reproduce this result, we
need the
gaussian form of a string wave--packet, which was constructed in
Ref.\cite{noi3} as an explicit solution of the functional Schr\"odinger
equation for loops. For our purposes Eq.(\ref{ondadue}) is quite
appropriate: rather than Fourier expanding the string coordinates
$x^\mu(s)$ and decomposing the functional wave equation into an infinite set of
ordinary differential equations, we insist in maintaining the
``wholeness'' of the string and consider {\it exact
solutions in loop space,} or adopt a {\it minisuperspace}
approximation quantizing only one, or few oscillation modes,
freezing all the other (infinite) ones. In the first case, it
is possible to get exact ``free'' solutions, such as the {\it
plane wave}
\begin{equation}
\Psi[\sigma\ ;A]\propto \exp
i\left\{{\cal E}\, A-\oint_C x^\mu\, dx^\nu
P_{\mu\nu}(x)\right\}\label{piana}
\end{equation}
which is a simultaneous
eigenstate of both the area momentum and Hamiltonian
operators. This wave functional describes a completely
unlocalized state: any loop shape is equally likely to occur
and therefore
the string has no definite shape. A physical state of
definite shape is obtained
by superposition of  the fundamental plane wave solutions
(\ref{piana}). The quantum analogue of a classical string is
the {\it Gaussian wave packet:}
\begin{eqnarray}
\Psi[\sigma,A]=&&\left[{1\over
2\pi(\Delta\sigma)^2}\right]^{3/4}\left( 1+{iA\over
m^2(\Delta\sigma)^2}\right)^{-3/2}\times\nonumber\\
&&\times\exp\left\{ {1\over 1+ (iA/ m^2(\Delta\sigma)^2)}\left[
-{\sigma^2\over 4(\Delta\sigma)^2}+i\oint_C x^\mu dx^\nu
P_{\mu\nu}(x)\right.\right.\nonumber\\
&&\qquad\left.\left. -{iA\over 4m^2l_C}\oint_C
d\mu(s)P_{\mu\nu}P^{\mu\nu} \right] \right\}\label{gauss}
\end{eqnarray}
where the width $\Delta\sigma$ of the packet at $A=0$
represents
the {\it area uncertainty}. From the solution (\ref{gauss})
one can derive some interesting results. First, we note that
the center of the
wave packet drifts through loop space according to the
stationary phase principle
\begin{equation}
\sigma^{\mu\nu}[C]-{A\over m^2}P^{\mu\nu}[C]=0 \end{equation}
and spreads as $A$ increases
\begin{equation}
\Delta\sigma\longrightarrow
\Delta\sigma(A)=\Delta\sigma\left(1+ {A^2\over 4m^4
(\Delta\sigma)^4}\right)^{1/2}   .\end{equation}
 Thus, as the areal time $A$ increases, the string ``decays'',
in the sense that it loses
its sharply defined initial shape, but in a way which is
controlled by the
{\it loop space
uncertainty principle}
\begin{equation}
{1\over 2}\Delta \sigma^{\mu\nu}[C] \Delta
P_{\mu\nu}[C]\gtrsim 1 \ , \qquad\hbox{in natural units}
\end{equation}
involving the uncertainty in the loop area, and the rate of
area
variation.\\
The central result that follows from the above equations, is
that the classical world--sheet of a string, a smooth
manifold of topological dimension two, turns
into a  {\it fractal object} with {\it Hausdorff dimension
three} as a consequence of the quantum areal fluctuations of
the string loop \cite{noi3}. With historical hindsight, this
result is not too surprising. Indeed, Abbott and Wise,
following an earlier insight by Feynman and Hibbs, have shown
in Ref. \cite{abbw} that the {\it quantum
trajectory} of a point-like particle is a fractal of
Hausdorff
dimension two. Accordingly, one would expect that the limiting
Hausdorff dimension of the world--sheet of a string increases by one unit since
one is dealing
with a one parameter family of one--dimensional
quantum trajectories. Next, we try to quantify the transition
from the classical,
or smooth
phase, to the quantum, or fractal phase. Use of the Shape
Uncertainty
Principle, and of the explicit form of the loop wave--packet,
enables
us to identify the control parameter of the transition with 
the De 
Broglie area characteristic of the loop. As a matter of fact, 
the Gaussian wave packet (\ref{gauss})
allows us to extend the Abbott and Wise calculation to the 
string 
case. By introducing the analogue of the ``De Broglie
wavelength $\Lambda$''as the inverse modulus of the loop
momentum\footnote{$[\Lambda]=length^2$. Accordingly, the
loop wavelength is strictly given by $\sqrt\Lambda$.}
\begin{equation}
{1\over 2}P^{\mu\nu}P_{\mu\nu}=\Lambda^{-2} \end{equation}
one finds:\\
i) at {\it low resolution}, i.e., when the area uncertainty
$\Delta\sigma >> \Lambda $, the Hausdorff dimension matches
the topological value, i.e., $D_H=2$; \\
ii) at {\it high resolution}, i.e., when the area uncertainty
$\Delta\sigma << \Lambda $, the Hausdorff dimension increases
by one unit, i.e., $D_H=3$.\\
Hence, quantum string dynamics can be described in terms of a
fluctuating Riemannian two--surface only when the observing
apparatus
is characterized by a low resolution power. As smaller and 
smaller areas are approached, the graininess of the world--sheet 
becomes manifest. Then a sort of {\it de--compactification} 
occurs, in the sense that the thickness of 
the string history comes into play, and the ``world--surface'' is literally
{\it fuzzy} to the extent that its Hausdorff dimension can be anything 
between its topological value of two and its limiting fractal value of three.

\subsection{Superconductivity and Quantum Spacetime}

Quantum strings, or more generally branes of various kind, 
are currently 
viewed as the fundamental constituents of everything: not 
only every matter particle or gauge boson must be 
derived from the string vibration spectrum, but spacetime 
itself is built out of them.\\
If spacetime is no longer preassigned, then logical
consistency demands that a matrix representation of p--brane
dynamics cannot be formulated in any given background
spacetime. The exact mechanism by which M--Theory is supposed
to break this circularity is not known at present, but ``loop
quantum mechanics'' points to a possible resolution of that
paradox. If one wishes to discuss quantum strings on the same
footing with point--particles and other p--branes, then their
dynamics is best formulated in loop space rather than in
physical spacetime. As we have repeatedly stated throughout
this paper, our emphasis on string {\it shapes} rather than
on the string constituent points, represents a departure from
the canonical formulation and requires an appropriate choice
of dynamical variables, namely the string configuration
tensor and the areal time. Then, at the loop space level,
where each ``point'' is representative of a particular loop
configuration, our formulation is purely quantum mechanical,
and there is no reference to the background spacetime. At the
same time, the functional approach leads to a precise
interpretation of the fuzziness of the underlying quantum
spacetime in the following sense: when the resolution of the
detecting apparatus is smaller than a particle De Broglie
wavelength, then the particle quantum trajectory behaves as a
fractal curve of Hausdorff dimension two. Similarly we have
concluded on the basis of the ``shape uncertainty principle''
that the Hausdorff dimension of a quantum string world--sheet
is three, and that two distinct phases (smooth and fractal
phase) exist above and below the loop De Broglie area. Now,
if particle world--lines and string world--sheets behave as
fractal objects at small scales of distance, so does the
world--history of a generic p--brane including spacetime
itself \cite{nott}, and we are led to the general expectation
that a new kind of {\it fractal geometry} may provide an
effective dynamical arena for physical phenomena near the
string or Planck scale in the same way that a smooth
Riemannian geometry provides an effective dynamical arena for
physical phenomena at large distance scales.\\
Once committed to that point of view, one may naturally ask,
``what kind of  physical mechanism can be invoked in the
framework of loop quantum mechanics to account for the
transition from the fractal to the smooth geometric phase of
spacetime?''. A possible answer consists in the phenomenon of
{\it p--brane condensation}. In order to illustrate its
meaning, let us focus, once again, on string loops. Then, we
suggest that what we call ``classical spacetime'' emerges as a
condensate, or string vacuum similar to the ground state of a
superconductor. The large scale properties of such a state
are described by an ``effective Riemannian geometry''. At a
distance scale of order $(\alpha')^{1/2}$, the condensate
``evaporates'', and with it, the very notion of Riemannian
spacetime. What is left behind, is the fuzzy stuff of quantum
spacetime. \\
Clearly, the above scenario is rooted in the functional
quantum mechanics of string loops discussed in the previous
sections, but is best understood in terms of a model which
mimics the Ginzburg--Landau theory of superconductivity. Let
us recall once again that one of the main results of the
functional
approach to quantum
strings is that {\it it is possible to describe the evolution
of the system without giving up
reparametrization invariance}. In that approach,
the clock that times the evolution of a closed bosonic string
is the {\it internal area,}
i.e., the area measured in the string parameter space
$\displaystyle{A\equiv {1\over 2}\epsilon_{ab}\int_D d\xi^a\wedge d\xi^b}$,
of {\it any} surface subtended by the string loop. The choice of
such a surface is arbitrary, corresponding to the freedom of choosing the
initial instant of time, i.e., a fiducial reference area.
Then one can take advantage of this arbitrariness in
the following way. In particle field theory an arbitrary lapse of euclidean,
or Wick rotated, imaginary time between initial and final
field configurations is usually given the meaning of {\it inverse temperature}
\beq
i\Delta t\longrightarrow \tau\equiv {1\over \kappa_B T}
\eeq
and the resulting euclidean field theory provides a {\it finite temperature} 
statistical description of vacuum fluctuations. \\
Following the same procedure, we suggest to analytically 
extend  $A$ to imaginary values,
$iA\longrightarrow a$, on the assumption that the resulting
{\it finite area loop quantum mechanics} will provide a 
statistical description of the string vacuum 
fluctuations. This leads us to consider the following 
effective (euclidean) lagrangian of the Ginzburg--Landau type,
\beqa
&&L(\Psi, \Psi^*)=\Psi^*
{\partial\over\partial a }\Psi- {1\over 4 m^2}
\left(\oint_C dl(s)\right)^{-1}\oint_C dl(s)\bigg\vert
\left({\delta \over \delta\sigma^{\mu\nu}(s)}-ig
A_{\mu\nu}(x)\right)\Psi
\bigg\vert^2 +\nonumber\\&&\phantom{(\Psi, \Psi^*)}
-V(\vert\Psi\vert^2)-{1\over 2\cdot
3!}H^{\lambda\mu\nu}H_{\lambda\mu\nu}
\label{model}\\
&&V(\vert\Psi\vert^2)\equiv
\mu_0^2\left({a_c\over a}-1\right)\vert\Psi\vert^2
+{\lambda\over 4}\vert\Psi\vert^4\\
&&H_{\lambda\mu\nu}=\partial_{[\lambda}A_{\mu\nu]}
\quad .
\eeqa
Here, the string field is coupled to a Kalb--Ramond gauge potential
$A_{\mu\nu}(x)$ and $a_c$ represents a {\it critical loop area} such that,
when $a\le a_c$ the potential energy is minimized by the {\it ordinary vacuum}
$\Psi[C]=0$, while for $a> a_c$ strings condense into a {\it superconducting
vacuum}. In other words,
\beq
\vert \Psi[C]\vert^2= \cases{0 &if $ia\le a_c$\cr
{\mu_0^2\over\lambda}\left(1-a_c/ a\right) &if  $a> a_c$}
\eeq
Evidently, we are thinking of the string condensate as the
large scale, background  spacetime. On the other hand, as one
approaches distances of the order $(\alpha')^{1/2}$ strings undergo
more and more shape--shifting transitions which destroy the
long range correlation of the string condensate. As we have
discussed earlier, this signals the transition from the
smooth to the fractal phase of the {\it string world--surface}.
On the other hand, the quantum mechanical approach
discussed in this paper is in no way restricted to string--like objects.
In principle, it can be extended to any quantum p--brane, and we
anticipate that the limiting value of the corresponding
fractal dimension would be $D_H=p+2$. Then, if the above over all
picture is correct, spacetime fuzziness acquires a well defined meaning.
Far from being a smooth, four--dimensional manifold
assigned ``ab initio'', spacetime is, rather, a ``process in the
making'', showing an ever changing fractal structure which responds
dynamically to the resolving power of the detecting apparatus.
At a distance scale of the order of Planck's length, i.e., when
\beq
a_c=G_N
\label{newt}
\eeq
then the {\it whole of spacetime boils over} and no trace is
left of the large scale condensate of either strings or p--branes. \\
As a final remark, it is interesting to note that since the
original paper by A.D.Sakarov about gravity as spacetime elasticity,
$G_N$ has been interpreted as a {\it phenomenological parameter}
describing the large scale properties of the gravitational vacuum.
Eq.(\ref{newt}) provides a deeper insight into the meaning of $G_N$ as the
{\it critical value} corresponding to the transition point between an
``elastic'' Riemannian--type condensate of extended objects and a
quantum phase which is just a Planckian foam of fractal objects.

\begin{table}
\caption{Loop Space functionals and boundary fields}
\begin{tabular}{cc}
\\
\medskip
$H[C]=( 4m^2l_C)^{-1}\oint_C d\mu(s)P_{\mu\nu}(s)
P^{\mu\nu}(s) $ & (Schild) Loop Hamiltonian\\
\medskip
$H(s)=( 4m^2l_C)^{-1}P_{\mu\nu}(s) P^{\mu\nu}(s) $ & (Schild)
String Hamiltonian\\
\medskip
$d\mu(s)\equiv \sqrt{x^{\prime\, 2}(s)}
x^{\prime\,\mu}\ , x^\prime\equiv dx^\mu/ds$ & loop invariant
measure\\ \medskip
$ l_C\equiv \oint_C d\mu(s)$
& loop proper length \\
\medskip
$P_{\mu\nu}(s)=m^2 \epsilon^{mn}\partial_m x_\mu \partial_n
x_\nu $ & area momentum density\\
\medskip
$ P_{\mu\nu}[C]\equiv l_C^{-1}\oint_C d\mu(s) P_{\mu\nu}(s)$
& loop momentum \\
\end{tabular}
\label{uno}
\end{table}

\begin{table}
\caption{ The Particle/String ``Dictionary''}
\begin{tabular}{cc}
\\
medskip
\underbar{Physical object}:
& massive point--particle $\longrightarrow $ non--vanishing
tension string.\\
\medskip
\underbar{Mathematical model} :&
point $P$ in $R^{(3)}$ $\longrightarrow $ space-like loop $C$
in $R^{(4)}$.\\
\medskip
\underbar{Topological meaning}:&
boundary (=endpoint) of a line $\longrightarrow $ boundary of
an open surface.\\
\medskip
\underbar{Coordinates}:&
$\{ x^1,x^2,x^3\}\longrightarrow$
area element $\sigma^{\mu\nu}[C]=\oint_C x^\mu dx^\nu$\\
\medskip
\underbar{Trajectory}:&
1--parameter family of points $\{\vec x(t)\} \longrightarrow
\{x^\mu(s;A)\}$ 1--parameter family of loops.\\
\medskip
\underbar{Evolution parameter}:&
``time'' $t$ $\longrightarrow $ area $A$ of the string
manifold. \\ \medskip
\underbar{Translations generators}:&
$\hbox{spatial shifts:}\quad
{\partial\over \partial x^i} \longrightarrow \quad
\hbox{shape deformations:}
{\delta \over
\delta\sigma^{\mu\nu}(s)}$\\
\medskip
\underbar{Evolution generator}:&
$\hbox{time shifs:}\quad
{\partial\over \partial t}\longrightarrow\quad\hbox{ proper
area variations:}\quad
{\partial\over \partial A}$\\
\medskip
\underbar{Topological dimension}:&
particle trajectory $D=1$ $\longrightarrow $
string trajectory $D=2$.\\
\medskip
\underbar{Distance}:&
$(\vec x -\vec x_0)^2\longrightarrow (\sigma^{\mu\nu}[C]-
\sigma^{\mu\nu}[C_0])^2$\\
\medskip
\underbar{Linear Momentum}:&
rate of change of spatial position $\longrightarrow $ rate of
change of string shape.\\
\medskip
\underbar{Hamiltonian}:&
time conjugate canonical variable $\longrightarrow $ area
conjugate canonical variable.\\
\end{tabular}
\label{dictionary}
\end{table}

\end{document}